%% file: main.tex
\documentclass[12pt]{iopart}

\usepackage{graphicx,color} 
\usepackage[normalem]{ulem}
\usepackage[shortlabels]{enumitem}
           \setlist[enumerate, 1]{1\textsuperscript{o}}

\usepackage{tikz}
\usetikzlibrary{shapes.geometric, arrows}
\tikzstyle{omicron} = [rectangle, rounded corners, minimum width=3cm, minimum height=1cm,text centered, text width=3cm,draw=black, fill=red!30]
\tikzstyle{analysisReady} = [rectangle, rounded corners, text width=3cm,minimum width=3cm, minimum height=1cm,text centered, draw=black, fill=blue!30]
\tikzstyle{SNRFreqCut} = [rectangle, rounded corners, text width=3cm,minimum width=3cm, minimum height=1cm,text centered, draw=black, fill=blue!30]
\tikzstyle{DQVetoes} = [rectangle, rounded corners, text width=3cm,minimum width=3cm, minimum height=1cm,text centered, text width=3cm,draw=black, fill=blue!30]
\tikzstyle{OScans} = [rectangle, rounded corners, text width=3cm,minimum width=3cm, minimum height=1cm,text centered, text width=3cm,draw=black, fill=green!30]
\tikzstyle{Unlabeled} = [rectangle, rounded corners, text width=3cm,minimum width=3cm, minimum height=1cm,text centered, text width=3cm,draw=black, fill=yellow!30]
\tikzstyle{machine learning} = [diamond, minimum width=3cm, minimum height=1cm, text centered, draw=black, fill=orange!30]
\tikzstyle{WorkFlows} = [rectangle, rounded corners, text width=3cm,minimum width=3cm, minimum height=1cm,text centered, text width=3cm,draw=black, fill=cyan!30]
\tikzstyle{arrow} = [thick,->,>=stealth]

\newcommand\orangesout{\bgroup\markoverwith{\textcolor{orange}{\rule[0.5ex]{2pt}{0.4pt}}}\ULon}

\expandafter\let\csname equation*\endcsname\relax

\expandafter\let\csname endequation*\endcsname\relax

\usepackage{adjustbox}
\usepackage{amsmath}
\usepackage{amssymb}
\usepackage{mathtools}
\usepackage{amsthm}
\usepackage{algorithm}
\usepackage{algorithmic}
\usepackage[letterpaper,margin=1in]{geometry}
\usepackage[utf8]{inputenc}
\usepackage{epstopdf}
\usepackage{fancyvrb}
\usepackage{color}
\usepackage{float}

\usepackage{subcaption}
\graphicspath{{images/}}

\input{definitions}

\begin{document}

\title[Gravity Spy]{Gravity Spy: Integrating Advanced LIGO Detector Characterization, Machine Learning, and Citizen Science}

\author{M.~Zevin$^1$, S.~Coughlin$^1$, S.~Bahaadini$^2$, E.~Besler$^2$, N.~Rohani$^2$, S.~Allen$^3$, M.~Cabero$^4$, K.~Crowston$^5$, A.~Katsaggelos$^2$, S.~Larson$^1$, T.K.~Lee$^6$, C.~Lintott$^7$, T.~Littenberg$^8$, A.~Lundgren$^4$, C.~{\O}sterlund$^5$, J.~Smith$^9$, L.~Trouille$^3$, and V.~Kalogera$^1$}

\address{1 Center for Interdisciplinary Exploration and Research in Astrophysics (CIERA) and Dept. of Physics and Astronomy, Northwestern University, 2145 Sheridan Rd, Evanston, IL 60208, USA}
\address{2 Electrical Engineering and Computer Science, Northwestern University, Evanston, IL, 606201, USA}
\address{3 Adler Planetarium, Chicago, IL, 60605, USA}
\address{4 Max-Planck-Institut f\"{u}r Gravitationsphysik, Callinstrasse 38, D-30167 Hannover, Germany}
\address{5 School of Information Studies, Syracuse University, Syracuse, NY, 13210, USA}
\address{6 Department of Communication, University of Utah, Salt Lake City, UT 84112, USA}
\address{7 Department of Physics, University of Oxford, Oxford, United Kingdom}
\address{8 NASA/Marshall Space Flight Center, Huntsville, AL 35812, USA}
\address{9 Department of Physics, California State University Fullerton, Fullerton, CA, 92831, USA}
\eads{\mailto{zevin@u.northwestern.edu}, \mailto{scottcoughlin2014@u.northwestern.edu}}

\begin{abstract}
With the first direct detection of gravitational waves, the Advanced Laser Interferometer Gravitational-wave Observatory (LIGO) has initiated a new field of astronomy by providing an alternate means of sensing the universe. The extreme sensitivity required to make such detections is achieved through exquisite isolation of all sensitive components of LIGO from non-gravitational-wave disturbances. Nonetheless, LIGO is still susceptible to a variety of instrumental and environmental sources of noise that contaminate the data. Of particular concern are noise features known as \textit{glitches}, which are transient and non-Gaussian in their nature, and occur at a high enough rate so that accidental coincidence between the two LIGO detectors is non-negligible. Glitches come in a wide range of time-frequency-amplitude morphologies, with new morphologies appearing as the detector evolves. Since they can obscure or mimic true gravitational-wave signals, a robust characterization of glitches is paramount in the effort to achieve the gravitational-wave detection rates that are predicted by the design sensitivity of LIGO. This proves a daunting task for members of the LIGO Scientific Collaboration alone due to the sheer amount of data. In this paper we describe an innovative project that combines crowdsourcing with machine learning to aid in the challenging task of categorizing all of the glitches recorded by the LIGO detectors. Through the Zooniverse platform, we engage and recruit volunteers from the public to categorize images of time-frequency representations of glitches into pre-identified morphological classes and to discover new classes that appear as the detectors evolve. In addition, machine learning algorithms are used to categorize images after being trained on human-classified examples of the morphological classes. Leveraging the strengths of both classification methods, we create a combined method with the aim of improving the efficiency and accuracy of each individual classifier. The resulting classification and characterization should help LIGO scientists to identify causes of glitches and subsequently eliminate them from the data or the detector entirely, thereby improving the rate and accuracy of gravitational-wave observations. We demonstrate these methods using a small subset of data from LIGO's first observing run. 

\end{abstract}

\maketitle

\section{Introduction}

Following a major upgrade, advanced LIGO completed its first observing run (O1), which spanned from September 12, 2015 through January 19, 2016 \cite{aLIGOdetectors}. During this run, the LIGO detectors made the first direct detection of gravitational waves and the first observations of binary black hole coalescences~\cite{GW150914,GW151226,O1BBH}. With these detections, LIGO has initiated a new field of astronomy by providing an alternate means of sensing the universe. Over the coming years, the increased sensitivity of the LIGO detectors and additional interferometers joining the network of gravitational-wave observatories \cite{AdVirgo,KAGRA,observing} will further increase sensitivity to the gravitational universe. 

In order to detect gravitational waves, LIGO requires sensitivity to length fluctuations a thousandth the diameter of a proton in the 4-kilometer detector arms. In future observing runs this sensitivity will further increase; at design sensitivity LIGO aims to have the ability of detecting neutron star-neutron star mergers up to a distance of 200 Mpc \cite{aLIGO}. This high sensitivity is achieved through exquisite isolation of the lasers, mirrors, and all sensitive components of LIGO from non-gravitational-wave disturbances. However, LIGO detectors are still susceptible to non-cosmic disturbances that cause noticeable signals in the detectors. The effort to identify, characterize, and separate sources of noise from cosmic signals is paramount in achieving LIGO sensitivity goals \cite{DetcharGW150914}. 

Of particular concern are transient, non-Gaussian noise features known as \textit{glitches}. Glitches are instrumental or environmental in nature (caused by e.g. small ground motions, ringing of the test-mass suspension system at resonant frequencies, or fluctuations in the laser) and come in a wide variety of time-frequency-amplitude morphologies. These artifacts can produce false-positive results in gravitational-wave searches, reduce the significance of candidate gravitational-wave signals, corrupt data, bias astrophysical parameter estimation, and reduce the amount of analyzable data. The sensitivity of searches for unmodeled gravitational waves are especially limited by the high rate of glitches in LIGO \cite{DetcharGW150914,S6detchar}. In the 51.5 days of O1 alone, approximately $10^{6}$ glitches over a minimum signal-to-noise ratio (SNR) threshold of 6 were recorded. To maximize the gravitational-wave detection rate, the causes of glitches must be identified and fixed within the detectors (in the best case) or glitches must be removed from the data set. Identifying how many different glitches have a similar morphology is an important first step to this, allowing prioritizing by number and characteristics. Therefore, it is necessary to develop robust methods to identify and characterize glitches. 

Teaching computers to identify and morphologically classify glitches in detector data is a challenge. Only a small number of glitch classes have been understood to the level where they could be removed from the data with confidence. Attempts to use machine learning algorithms have shown promise in glitch classification endeavors \cite{old_glitch_classification,NN_glitch_classification,PCAT,PCAT2,diff_boosting_NN}, however these techniques do not yet capture the full range of glitch morphologies present in LIGO data. Though human ability to recognize patterns is a proven tool for such diverse classification endeavors, though the high volume of data that LIGO streams would easily overwhelm any small group of scientists. 

To address this challenge, we have developed \textit{Gravity Spy} - an interdisciplinary project that will leverage the strengths of both humans and computers to create a superior classifier of glitches in LIGO data. Gravity Spy addresses this task through the convergence of four science areas: gravitational physics, human-centered computing, machine learning, and citizen science. Specifically, the goal of the project is to leverage the advantages of citizen science along with those of machine- and human-learning techniques to design a socio-computational system with which to analyze and characterize LIGO glitches and improve the effectiveness of gravitational-wave searches. Gravity Spy also complements current glitch classification techniques, as it readily identifies new categories of glitches that arise as the detectors evolve, scales with an increasing number of unique glitch classes, and continually bolsters labeled set of prexisting classes. 

The Gravity Spy project couples human classification with machine learning models in a symbiotic relationship: volunteers provide large, labeled sets of known glitches to train machine learning algorithms and identify new glitch categories, while machine learning algorithms ``learn" from the volunteer classifications, rapidly classify the entire dataset of glitches, and guides how information is provided back to participants. The Gravity Spy project includes research on the human-centered computing aspects of this socio-computational system, as empirical testing of the human-computer interface leads to better project design and an enhanced performance of citizen science volunteers. Gravity Spy is implemented through Zooniverse.org, the leading online platform for citizen science, which has fielded a workable crowdsourcing model. Currently, over 1.5 million ``citizen scientists" work to provide analyses of scientific data on more than 40 projects \cite{zooniverse}. A beta version of Gravity Spy has already resulted in the identification of new glitch morphological classes, and shows promise for helping to improve LIGO data quality during upcoming observing runs. 

In this paper, we summarize the impact of glitches on LIGO data analysis and current efforts to mitigate their effects (Section~\ref{sec:detchar}). We then discuss the Gravity Spy project in full (Section~\ref{sec:gs}), highlighting in particular data preparation for the project (3.1), the citizen science interface (3.2), machine learning algorithms used for image classification and crowdsourcing classifiers (3.3), and social science experiments for the socio-computational system (3.4). Next we discuss preliminary results of the project using data from the first LIGO observing run (Section~\ref{sec:results}). Lastly, we comment on future prospects for the Gravity Spy project and its role in LIGO detector characterization (Section~\ref{sec:future}).

\section{Characterization of transient noise in LIGO}\label{sec:detchar}

\subsection{Impact of Glitches on Gravitational-Wave Data Analysis}

Searches for transient gravitational-wave signals, especially those that are short duration, in LIGO's sensitive frequency band, and/or poorly modeled~\cite{DetcharGW150914}, are highly susceptible to glitches in the data. One method for mitigating the impact of glitches is the requirement of coincidence between the LIGO observatories, which are located in Hanford, Washington and Livingston, Louisiana. Gravitational waves would appear in both detectors separated in time by less than or equal to the light travel time between the observatories. If a signal appears in only one observatory during this time window, it is rejected. Despite this requirement, glitches occur at a high enough rate that accidental coincidence between the two detectors is non-negligible. 

Glitches impact LIGO data analysis efforts in three critical ways. First, they increase the loudness of the background in gravitational-wave searches, which reduces the significance of candidate events. Even searches that utilize signal models to create discriminating signal statistics (e.g. compact binary coalescence searches~\cite{bruce2005,bruce2012}) are afflicted by glitch occurrences. Second, glitches impact the recovery of astrophysical parameters from a gravitational wave source~\cite{O1BBH,ninja2,PEGW150914}, since glitches that occur near the same time as a gravitational-wave signal reduce the SNR of the event and lead to broader uncertainties in parameter estimation. Finally, glitches reduce the amount of usable data. While data ``vetoes" can be constructed for times when glitches are known to occur, they eliminate the data available to be searched for astrophysical signals. Therefore, identifying the cause of glitches and eliminating the source of the glitch is much preferred to constructing such vetoes. The negative effects of glitches on data analysis make the identification and mitigation of glitches an essential part of the LIGO science effort. 

\begin{figure}
	\centering
    \begin{subfigure}[b]{\textwidth}
    	\captionsetup{font=small}
    	\caption{Blip glitch}
		\includegraphics[width=\textwidth]{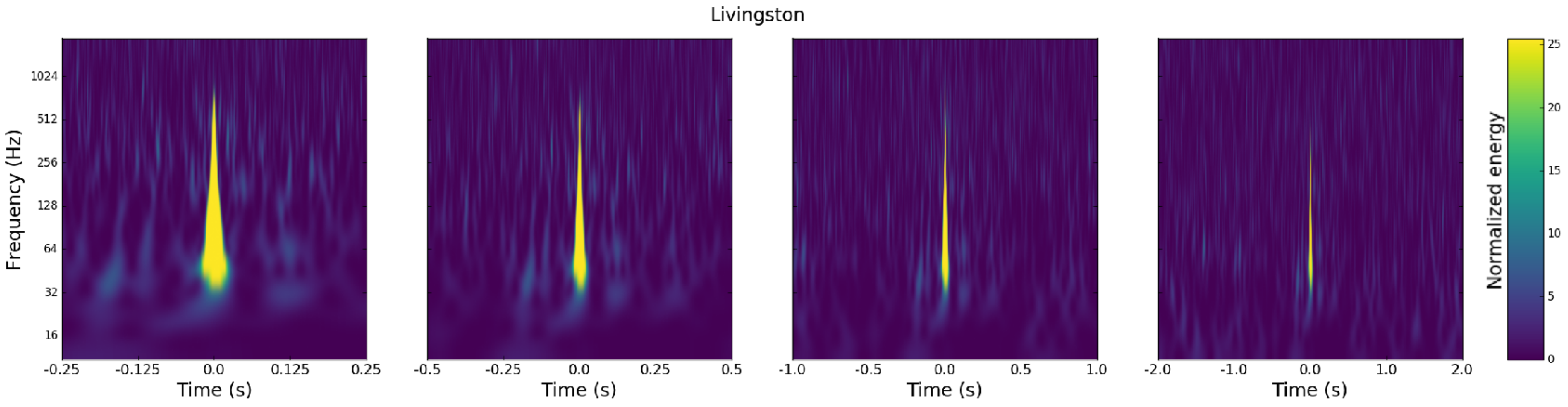}
        \label{fig:blip}
	\end{subfigure}
	\begin{subfigure}[b]{\textwidth}
    	\captionsetup{font=small}
        \caption{Whistle glitch}
		\includegraphics[width=\textwidth]{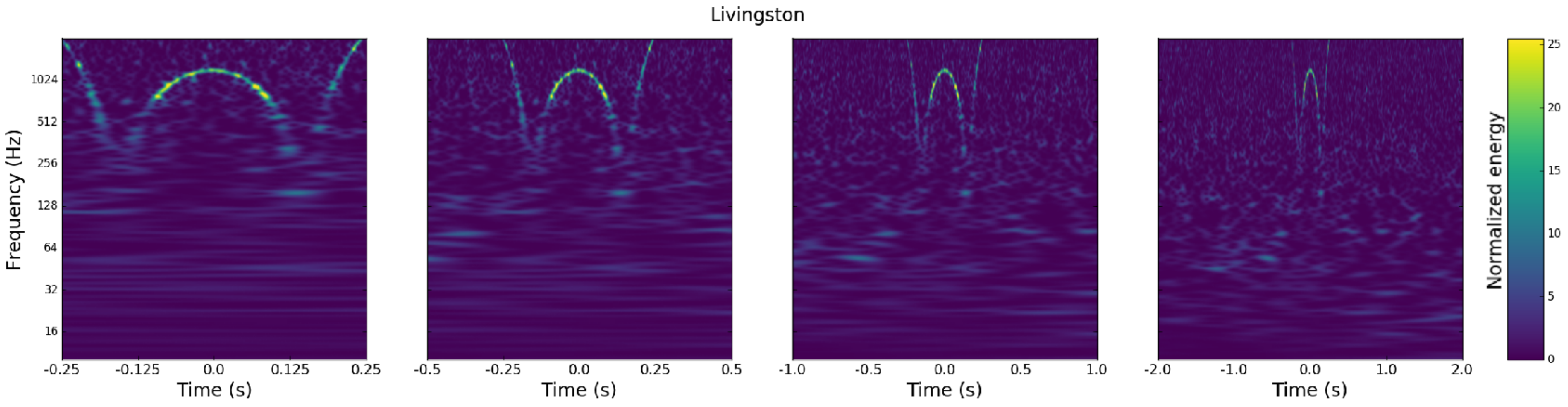}
        \label{fig:whistle}
	\end{subfigure}
    \captionsetup{font=small,skip=-15pt}
    \caption{Spectrogram representation of two example glitches, with color representing the `loudness' of the signal. Blips (a) are short glitches that usually appear in LIGO's gravitational-wave channel with a symmetric `teardrop' shape in time-frequency. Blips are the single most important class of glitches in LIGO \cite{DetcharGW150914}, as they appear in both Hanford and Livingston detectors and are the most stringent limit on LIGO's ability to detect binary black hole merger signals \cite{O1BBH}. No clear correlation to any auxiliary channel has yet been identified. Whistles (b), also known as radio frequency beat notes, usually appear in time-frequency plots with a characteristic `W' or `V' shape. Whistles are caused by radio signals at megahertz frequencies that beat with the LIGO Voltage Controlled Oscillators \cite{nutall_voltage_control}. These types of images are what volunteers in the Gravity Spy project classify, and what the associated machine learning algorithms use for training.}
\label{fig:common_glitches}
\end{figure}

\subsection{Identifying Glitches}

Several categories of glitches have been identified by the LIGO Scientific Collaboration (LSC), grouped by common origin and/or similar morphological characteristics \cite{PCAT,PCAT2,diff_boosting_NN}. Some of these categories have known causes, while others have causes yet to be identified. For example, two common morphological classes of glitches are shown in Figure \ref{fig:common_glitches}. Blip glitches (\ref{fig:blip}) are caused by unknown processes, whereas whistle glitches (\ref{fig:whistle})  are caused by radio signals at megahertz frequencies that beat with  Voltage Controlled Oscillators in the interferometer control system \cite{nutall_voltage_control}. 

Techniques have been developed to identify and categorize some categories of glitches automatically. Identification algorithms search for excess power in the time-frequency space of LIGO strain data and in hundreds of auxiliary channels, which are insensitive to gravitational waves and monitor the many instrumental and environmental factors potentially affecting the detectors. In addition to identifying a glitch, these algorithms parameterize glitches according to their time, frequency, SNR, and duration, among other parameters~\cite{omicron_alg,omicron}. Current approaches also search for statistical correlation between glitches in the gravitational-wave strain data channel and triggers in auxiliary channels~\cite{hveto, optimizing_vetoes, noise-coupling, used_percentage_veto}. However, due to the sheer volume of data, the LSC has not yet been able to filter through the millions of glitches to create a comprehensive categorization.

\subsection{Mitigating Glitches}

Having identified a glitch, the goal is to eliminate it from the detector. If the root cause of a glitch cannot be determined or its source cannot be fixed, information from glitch identification algorithms can be used to create data vetoes. Such vetoes improve gravitational-wave searches by removing times strongly affected by noise transients. 

Even these efforts, however, suffer from problems stemming from the very large number of glitches and their variety of morphologies. First, automated glitch classification algorithms have been unable to capture the varied morphological characteristics of all unique classes of glitches. In addition, certain types of glitches come and go over the course of an observing run, making their discovery challenging even for members of the LIGO science team. Finally, the software which implements  data quality vetoes would benefit from being fed information from specific categories of glitches instead of entire batches of glitches. This specificity would improve the ability to identify potential auxiliary channels that correlate with certain glitch morphologies, which in turn would contribute to identifying their source.

\section{Gravity Spy Project}\label{sec:gs}

The data challenges faced by LIGO are not unique. The increasingly large datasets that permeate every realm of modern science require new and innovative techniques for analysis \cite{data_scientific_discovery}. In astronomy, individual researchers have traditionally analyzed images of astronomical objects themselves; however, the digital surveys of today image hundreds of millions of objects, making the previous paradigm impractical. The acceleration in data acquisition has not been matched by an increase in human capacity to turn data into knowledge. 

Crowdsourcing data to volunteer citizen scientists offers one solution to this problem. Early efforts, such as NASA's Clickworkers, demonstrated the utility of crowdsourcing data to volunteers and the innate desire that the public has to contribute to scientific research \cite{clickworkers}. Another early astronomical project, Stardust@Home \cite{stardust}, led to the development of a general set of tools for citizen science projects known as BOSSA (now pyBOSSA\footnote{pybossa.com}). The highly successful Galaxy Zoo (e.g. \cite{galaxyzoo_lintott,galaxyzoo_galloway}) and Zooniverse projects (e.g. \cite{zooniverse_smith,zooniverse_kendrew,zooniverse_hennon,zooniverse_geach}) have demonstrated that it is possible to recruit hundreds of thousands of volunteers to make an authentic contribution to data analysis. To date, Zooniverse users have contributed to more than 100 peer-reviewed publications across a broad range of scientific disciplines. 

Glitch classification and characterization in LIGO currently utilizes human inspection, and therefore fits naturally into a citizen science framework. However, as scientific endeavors such as LIGO and future astronomical sky surveys become more data intensive, new methodologies must be explored for utilizing citizen scientists in data analysis. The Large Synoptic Survey Telescope (LSST), for example, will image tens of billions of galaxies \cite{LSST}, which is orders of magnitude more data than even the most successful citizen science projects can analyze. Supervised machine learning has proven to be a useful tool in projects which require a systematic analysis of substantial datasets such as these. However, these algorithms require a large, labeled dataset for training and struggle to identify new morphological categories as they appear. 

The data challenges faced in astronomy and other sciences today require a new generation of intelligent citizen science projects that are smarter about allocating tasks and more sophisticated in combining human and machine classification. This provides a two-way path to developing better machine learning algorithms and, for the first time with Gravity Spy, better human classifiers as well. Gravity Spy facilitates a symbiotic relationship between humans and computers, leveraging human pattern recognition skills as a tool for image recognition and machine learning as a tool for systematic analysis of large datasets. Citizen scientists analyze glitches from the LIGO data stream via human classification interfaces known as \textit{workflows}, providing labeled morphological classes as training data for machine learning algorithms. Trained machine learning algorithms classify the LIGO glitches data in full, determining confidence scores in each classification and feeding the most questionable glitches back to the citizen scientists for further analysis. 

\begin{figure}[t]
\centering
\includegraphics[width=\textwidth]{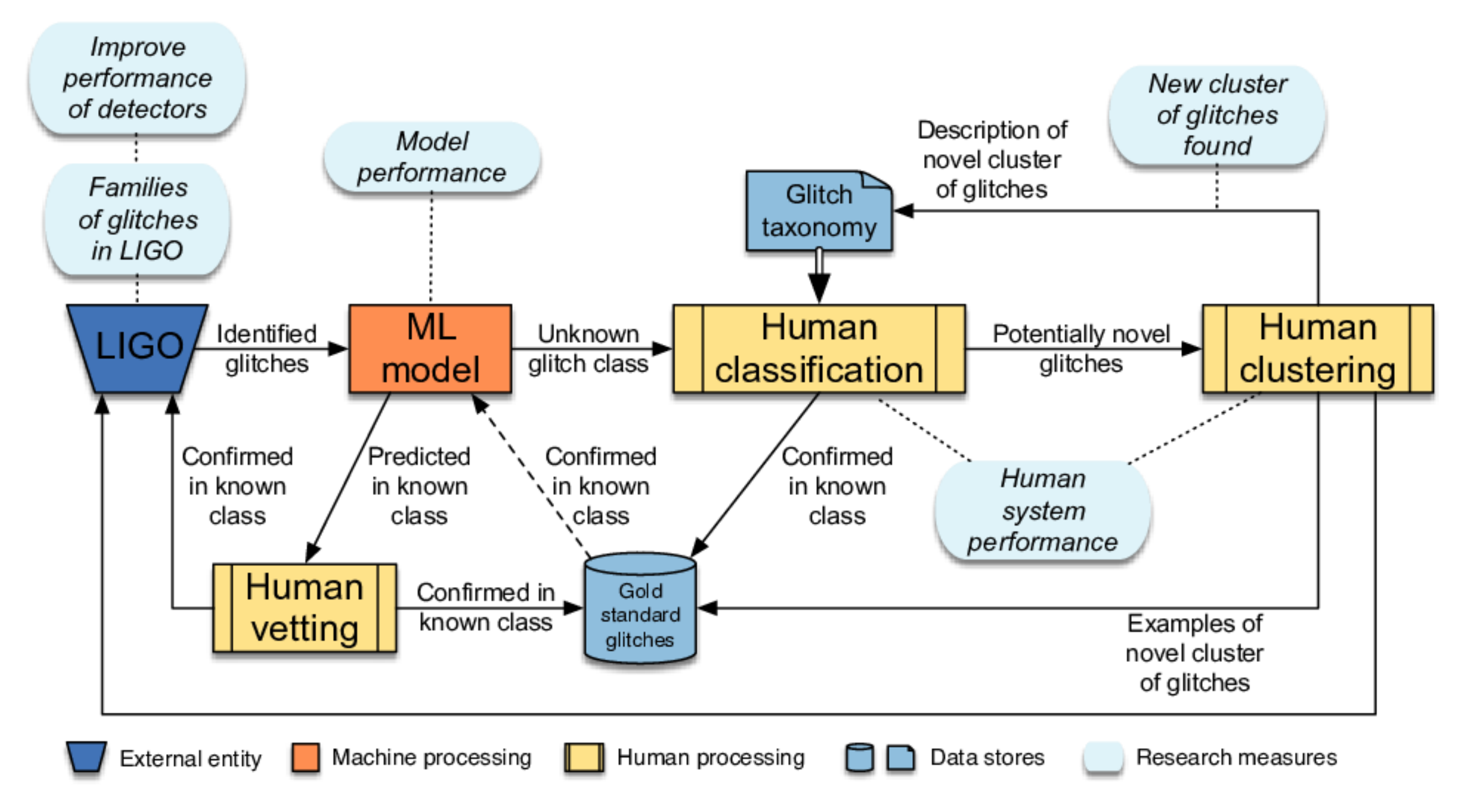}
\captionsetup{font=small,skip=-10pt}
\caption{Gravity Spy system architecture, and overall data flow through the interconnected, interdisciplinary components of the project. }
\label{fig:bigblockdiagram}
\end{figure}

A further innovation is that machine-analyzed glitches will guide training of new volunteers. As part of the Gravity Spy system, images classified by experts, known as `gold standard' images, are integrated into the user workflows. Individual user performance is analyzed by comparing that user's classifications with such gold standard images. This form of user analysis expedites the retirement of glitches and the growth of machine learning training sets (see Section \ref{subsec:crowdsource_classifier} for more details). Figure \ref{fig:bigblockdiagram} shows the interconnected components of the Gravity Spy project, and the movement of glitches through the project. 

Developing the next generation of citizen science projects requires significant advances in our understanding of human-centered computing. Studies of such projects have begun to answer important human-centered computing design questions \cite{human-centered_computing}, such as what kinds of tasks can non-experts perform reliably? What factors motivate participants? How do participants learn to perform the task or learn about the underlying science? Gravity Spy provides a platform to explore these questions more systematically, asking participants not only to apply existing scientific knowledge, but also to generate new knowledge (in this case, new categories of glitches). This setting allows the exploration of additional questions, such as how to support not just individual citizen scientists but teams working together, and what organizational structures are most appropriate?

Finally, Gravity Spy addresses the pressing need to understand the development of socio-computational systems that merge the distinctive strengths of computers (i.e. the ability to process large amounts of data systematically) and the humans (i.e. the ability to see patterns and spot discrepancies) \cite{socio-computational_systems,gaming,crowds_wikipedia,handbook_collective_intelligence}. Knowledge of how to use human-coded data to improve machine learning (e.g. by applying an active learning approach) is fairly well developed, though there are still opportunities to study the human-interface aspects of the process. In contrast, we still know little about how to use machine-analyzed data to improve human performance and thus how we best leverage human learning and machine learning in a joint effort. 

\subsection{Data Preparation}\label{sec:dataprep}

Data preparation for the Gravity Spy project (i.e. the link from LIGO to the rest of the project in Figure \ref{fig:bigblockdiagram}) presented three critical challenges: 
\begin{enumerate}[1.]
\item
Given that during O1 alone there were more than $10^{6}$ glitch triggers identified by the Omicron transient search algorithm \cite{omicron_alg,omicron}, it is crucial to determine which glitches were best fit for volunteer classification and most useful for LIGO detector characterization and data analysis
\item
Deciding the proper presentation of the morphologically-diverse zoo of glitches to both volunteers and machine learning algorithms
\item
Since there is no complete catalog of glitch categories that appeared during O1, the preparation of a training set needed to develop organically from various sources associated with the project
\end{enumerate}

\subsubsection{Data Selection}\label{sec:DataSel}

In order to tackle the first challenge, we only use glitches that satisfied the following criteria. First, the glitch occurs while the detector is in \textit{lock} and in observing mode, meaning the state of the detector was adequate enough to be searching for gravitational waves and ready for data analysis. For O1 glitches, we also neglected times that were flagged for poor data quality (DQ), though depending on the latency at which such flags are raised in future observing runs this cut may not be applied when feeding data into the system. Second, we neglect glitches where the SNR reported by the Omicron search pipeline is below 7.5, as glitches below this threshold prove to be exceedingly difficult to classify by eye. Third, the peak frequency of the glitch falls between 10 Hz and 2048 Hz. These choices are motivated by our goal to analyze and understand glitches that have the largest impact on the gravitational-wave searches: low-SNR glitches are less detrimental to searches, and this frequency range aligns well with LIGO's most sensitive frequency band and the frequency range expected for compact binary coalescence gravitational-wave events. In addition, as the Gravity Spy pipeline was first run after the conclusion of O1, we had the benefit of being able to apply the same DQ vetoes \cite{DetcharGW150914,hveto} to the data as were applied during astrophysical searches. Again, this was in order to analyze the glitches that have the largest impact on gravitational-wave searches.

The gravitational-wave events GW150914 \cite{GW150914} and GW151226 \cite{GW151226} and gravitational-wave trigger LVT151012 \cite{O1BBH} are not included in the Gravity Spy dataset. Hardware injections \cite{hardware_injections} are included, and constitute most of the subjects in the `Chirp' glitch class. However, in future observing runs potential gravitational-wave signals will not necessarily be redacted, as new images will be added to the project before the results of gravitational-wave searches are available. To ensure astrophysical claims cannot be made by non-LSC users, GPS times are replaced by a random, unique ID for each image in the Gravity Spy system. Therefore, potential astrophysical signals will be indistinguishable from hardware injections in the detectors, and users will have no knowledge of when a particular trigger was recorded.

\subsubsection{Omega Scans}\label{sec:OmScan}

We met the second challenge by representing these glitches with Omega Scans \cite{OmegaScan}. Omega Scans originated as a pipeline for the detection of gravitational wave transients, and are similar to spectrograms in that they represent glitches in time-frequency-energy space. They are also excellent at visualizing glitches that may cause problems in gravitational-wave searches. Omega Scans represent a generic signal as a combination of sine-Gaussians. The main utility of Omega Scans is an unmodeled SNR calculation with the template for a signal defined by its `Q' value, where Q is the quality factor of a sine-Gaussian waveform. In practice, this template signal consists of a time-frequency tiling. Like all template searches, an Omega Scan searches over a range of Q templates (i.e. time-frequency tilings) and identifies the template that gives the loudest SNR value. After identifying the Q template that provides the loudest value, the most significant tile for that Q template is identified and a spectrogram is generated. The color scale of the image is the Normalized Energy, which is directly related to the SNR of a tile and defined as the square of a given tile's Q transform magnitude divided by the mean squared magnitude in the presence of stationary white noise: 
\begin{equation}
Z = \frac{|X|^{2}}{\langle|X|^{2}\rangle}
\end{equation}
\noindent
where $Z$ is the normalized energy and $|X|$ is the Q transform magnitude of a tile \cite{OmegaScan}. 

As shown in Figure \ref{fig:common_glitches}, each image has the glitch fixed at the center of the Omega Scan, and each glitch is visualized using four different time windows ($\pm$ 0.25, 0.5, 1.0, and 2.0 seconds) to accommodate both long-duration and short-duration glitches. Human volunteers and machine learning algorithms are presented all four time durations of each glitch for classification purposes. 

\subsubsection{Training Set}\label{subsec:training_sets}

The final challenge was the construction of a large and accurately-labeled set of LIGO glitches. The generation of such \textit{training sets} is one of the most difficult components of supervised machine learning, and necessary to properly train classification algorithms. The past attempts to compile glitches into morphological classes using computer algorithms (e.g. \cite{old_glitch_classification,NN_glitch_classification,PCAT,PCAT2,diff_boosting_NN}) often rely solely on raw data or metadata from search pipelines rather than by-eye classification. In addition, new glitch morphologies that appeared during the first observing run of LIGO were not analyzed nor categorized to the level of pre-existing glitches.

\begin{table}[b]
\centering
\caption{Breakdown of morphological categories in the Gravity Spy training set, indicating the component of each class that comes from Livingston detector data and Hanford detector data.}
\includegraphics[width=\textwidth]{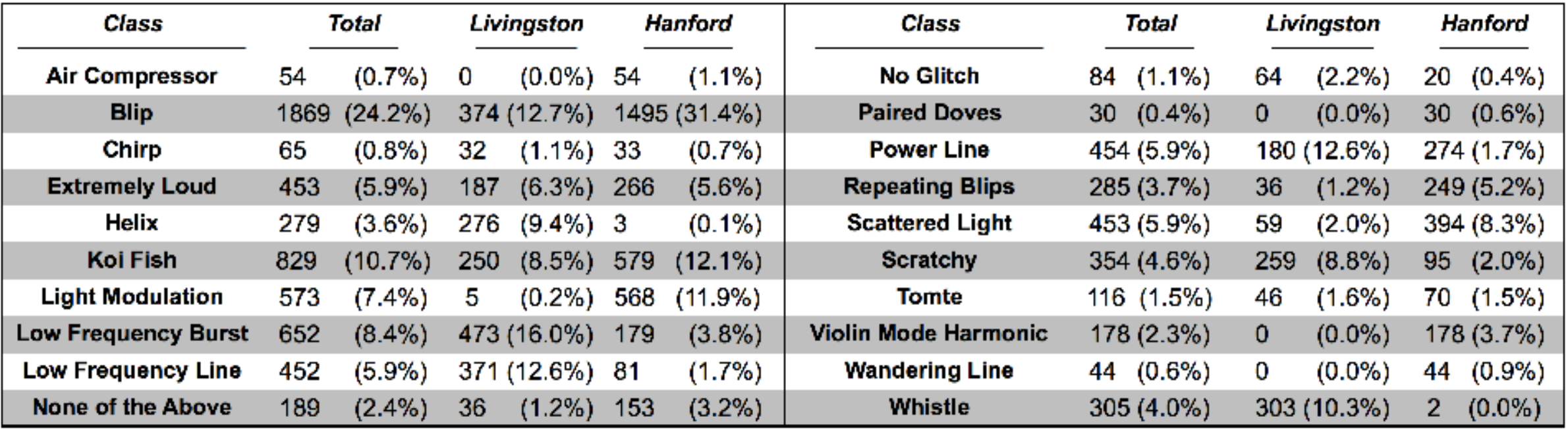}
\label{tbl:class_numbers}
\end{table}

A training set of glitches from O1 was generated for the Gravity Spy project by observing large quantities of Omega Scans and categorizing the images by morphology, with the aid of simple machine learning algorithms. First, consultation with LIGO detector characterization experts helped identify a few prominent and documented classes of glitches. Omega Scans of all LIGO Omicron triggers within the frequency and SNR cuts specified above were generated, which reduced the dataset to about $10^{5}$ glitches for the entirety of O1. We proceeded by classifying glitches from this set into preexisting categories based on the morphology of the glitch in its Omega Scan, and new categories of glitches were identified and accumulated in the process. Due to the similar morphological characteristics of many glitch classes, this process took multiple iterations to assure reliability in the class differentiation. Nonetheless, this tactic only accumulated $\sim100$ glitches per class. 

This small set of human-identified glitches was used to train preliminary machine learning algorithms to classify the remainder of the glitch dataset. Though such algorithms  only achieved classification accuracy of $\sim80\%-90\%$, they were useful in differentiating the unlabeled dataset into morphologically-similar classes, thus fostering an easier by-eye classification process. As will be described in Section \ref{subsec:beta}, during the beta-testing of the project, two new classes were also identified and characterized by Gravity Spy volunteers. Additional training data for these new classes was identified using the same methods described above. 

In total, a labeled training set of $7718$ glitches was built from both the Livingston and Hanford detectors for the preliminary machine learning analyses presented in this paper. These glitches are grouped into 20 classes, with exact proportions shown in Table \ref{tbl:class_numbers}. Given that each glitch is imaged at a maximum duration of 4 seconds, this amounts to 8.58 hours (0.7\%) of O1 data \cite{O1BBH}.

\subsection{Citizen Science}

Once the Omega Scans of glitches are in the system, they are classified by Gravity Spy volunteers, who populate the human-classification unit of the system (green boxes in Figure \ref{fig:bigblockdiagram}). 

\subsubsection{User Interface}

\begin{figure}
\centering
\includegraphics[width=\textwidth]{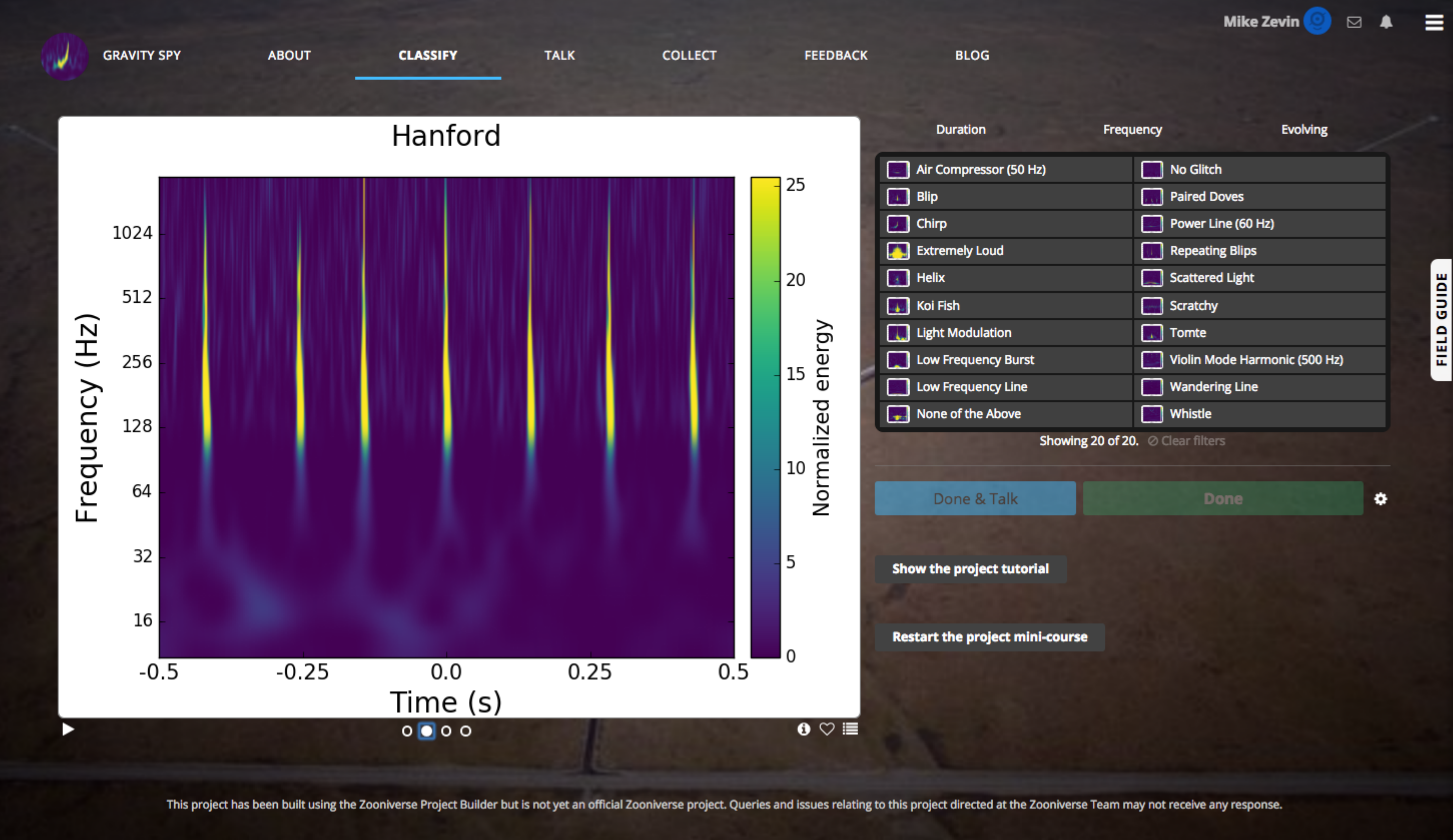}
\caption{Gravity Spy user interface. This image shows the \textit{Black Hole Merger} workflow (see Section \ref{sec:workflows}), with all 20 currently designated categories as options.}
\label{fig:user_interface}
\end{figure}

The user interface for Gravity Spy was created using the Zooniverse DIY Project Builder\footnote{Zooniverse.org/lab}, which enables anyone to build their own Zooniverse citizen science project for free through a set of easy-to-use, browser-based tools. The Gravity Spy classification interface containing the currently-known 20 glitch classes as options is shown in Figure \ref{fig:user_interface}. Example images of each glitch morphology can be found on the Gravity Spy website\footnote{gravityspy.org}. Through this interface, volunteers are shown individual Omega Scans of glitches to classify into one of the categories. Volunteers have the ability to cycle through multiple renderings of a given glitch over differing time durations, enabling a volunteer to visualize both long-duration and short-duration glitches. After classifying a glitch, the volunteer has the option of moving on to further classifications, or posting the glitch to `Talk', which is the Zooniverse discussion forum that provides a basis for interaction between Zooniverse volunteers and Gravity Spy project scientists.

Clicking on any glitch morphology option provides basic written information about that class to the volunteer along with multiple example images belonging to that glitch class. In addition, this dialog contains images of glitch morphologies that are often confused with that class, providing a simple means of changing classification choice to similar glitches if the volunteer misidentified the image initially. Alternatively, users can narrow down glitch options by filtering based on how long the glitch persists (\textit{duration}), the characteristic frequency of the glitch (\textit{frequency}), and whether the glitch is evolving in time (\textit{evolving}). Further information regarding each glitch class can be found in the \textit{Field Guide} (visible on the right side of Figure \ref{fig:user_interface}) . 

If a glitch does not fit into any of the predefined categories, a user can classify it as ``None of the Above". In doing so, a volunteer is asked follow-up questions describing the morphology of the glitch (i.e. information about its duration, frequency, and time evolution). By this process and through user activity on Talk, new classes of glitches can be identified and integrated into the Gravity Spy project. This allows the Gravity Spy glitches classes to evolve and follow changes in the glitch types that occur in the LIGO detectors. 

\subsubsection{Volunteer Training}

A key question in citizen science is how reliably volunteers perform the classification task, known as \textit{results quality}. Zooniverse’s approach to citizen science directly addresses this question and has led to an established track record of producing quality data for use by the wider scientific community and publications across the disciplines. By embedding training within the interface and creating consensus results based on numerous classifications for each image, Zooniverse projects help to make a disparate crowd of volunteers produce reliable results \cite{zooniverse}.

As with other Zooniverse projects, Gravity Spy begins with a brief tutorial, explaining the project's goals, how to interpret the spectograms, and how to use the classification interface. The Field Guide and additional content pages describe properties of each glitch class and the LIGO project in more detail. 

Research on learning suggests that an effective way to train humans to perform image classification tasks is to provide them with exemplary images from which to learn \cite{kim_murphy, Kulatunga}. Accordingly, as in other citizen science projects, the Gravity Spy classification interface shows the volunteers example images of all the glitch classes to guide the choice. 

An advance over the current state of the art citizen science project is that Gravity Spy uses machine learning results to train the human volunteers more systematically. Specifically, the system moves new volunteers through a sequence of levels in which they are presented with an increasing number glitches classes and sophistication of features within the classification interface, intended to improve their ability to classify glitches \cite{roads_mozer}. Essentially, the system is `tutoring' volunteers, but rather than simply taking images from a predefined set of training materials, it identifies novel images in need of classification that should still help beginners to learn. 

Upon joining a project, a volunteer is presented glitches that have been classified by the machine learning models as likely belonging to only one of two very distinctive classes. For each glitch, volunteers are asked to annotate it as being an instance of one of the two classes or “None of the Above” (a reduced version of the interface shown in Figure 4). These exemplary images help the volunteer to learn how to identify this subset of glitch classes. Once volunteers are reliably classifying these two initial classes, additional classes are introduced.  

In the current implementation, volunteers also classify gold standard images, which in practice are a subset of the full machine learning training set. After classifying a gold standard image, the volunteer immediately receives feedback as to whether their classification agrees with the expert classification. Initially, 40\% of images presented to beginning volunteers are gold standard, and this frequency dynamically decreases as a volunteer classifies gold standard images correctly.

As volunteers progress through the training regimen, they are presented with more classes that the machine learning model has classified with high confidence. The classifications during this training period contribute to the project by verifying the high-confident, yet imperfect, machine learning results. In addition to training the volunteers in recognizing members of more glitch classes, the levels are expected to motivate users by appealing to their sense of accomplishment. 


Once the user has completed multiple rounds of training on a subset of glitch classes with high machine learning confidence scores, they are considered fully qualified and will be given glitches to classify at varying levels of machine learning confidence in all known classes or even glitches for which the machine learning has no good classification, thus further contributing to the identification of new glitch categories. Since the system tracks each volunteer’s reliability, it can also assign tasks based on the capabilities of each volunteer. 

\subsubsection{Workflows}\label{sec:workflows}

\begin{figure}
\centering
\makebox[\textwidth][c]{\includegraphics[width=1.1\textwidth]{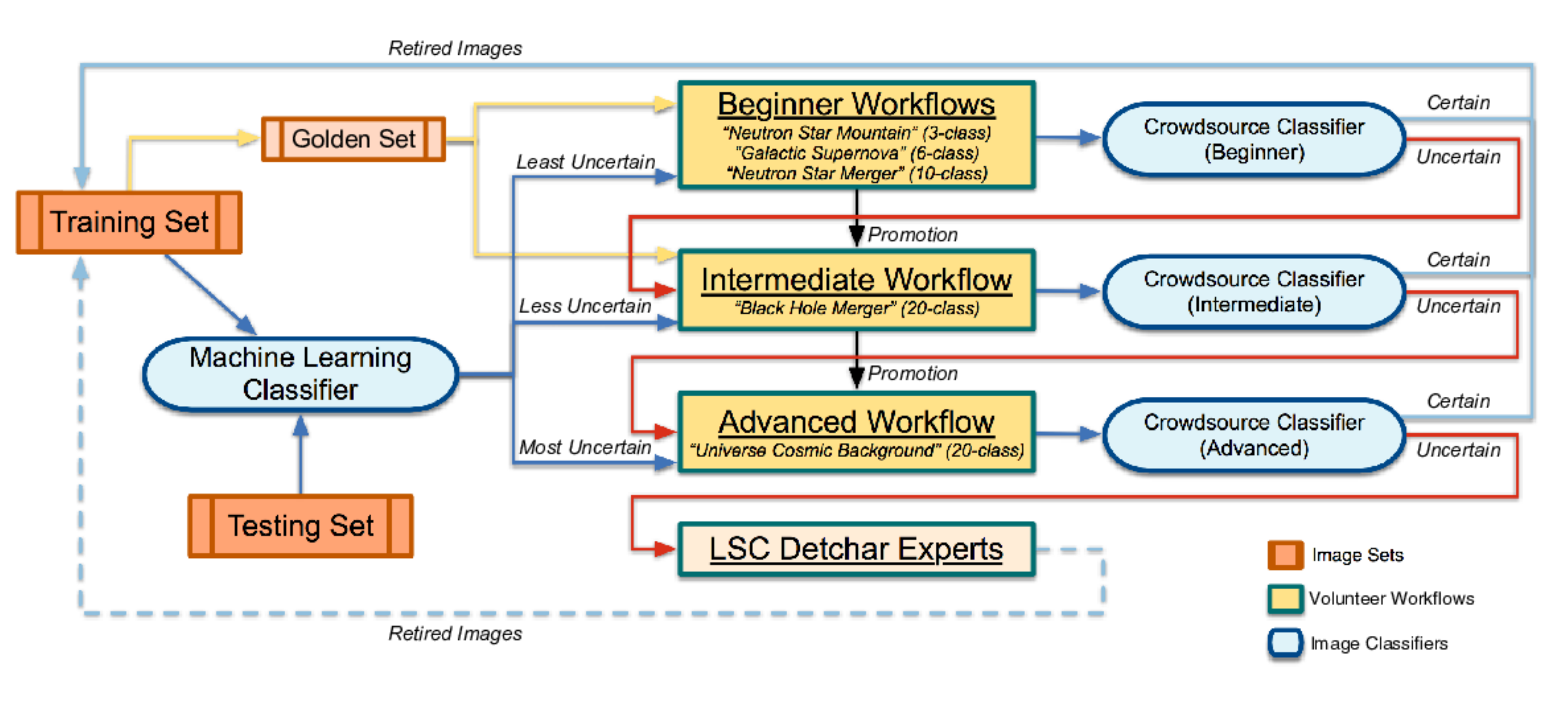}}
\caption{Movement of images and volunteers through the Gravity Spy project. Green boxes represent the multiple workflows within the project (including the images which are forwarded to experts within the LSC), blue boxes represent the machine learning and crowdsourcing image classifiers, and orange boxes represent the full sets of images, which are designated either as training or testing images (the `golden set' is the subset of the training set which is used to train volunteers). Note that there are multiple beginner workflows with an increasing number of glitch classes which volunteers progress through as they proceed through the training regimen.}
\label{fig:workflow_flowchart}
\end{figure}

\begin{figure}[t]
\centering
\includegraphics[width=\textwidth]{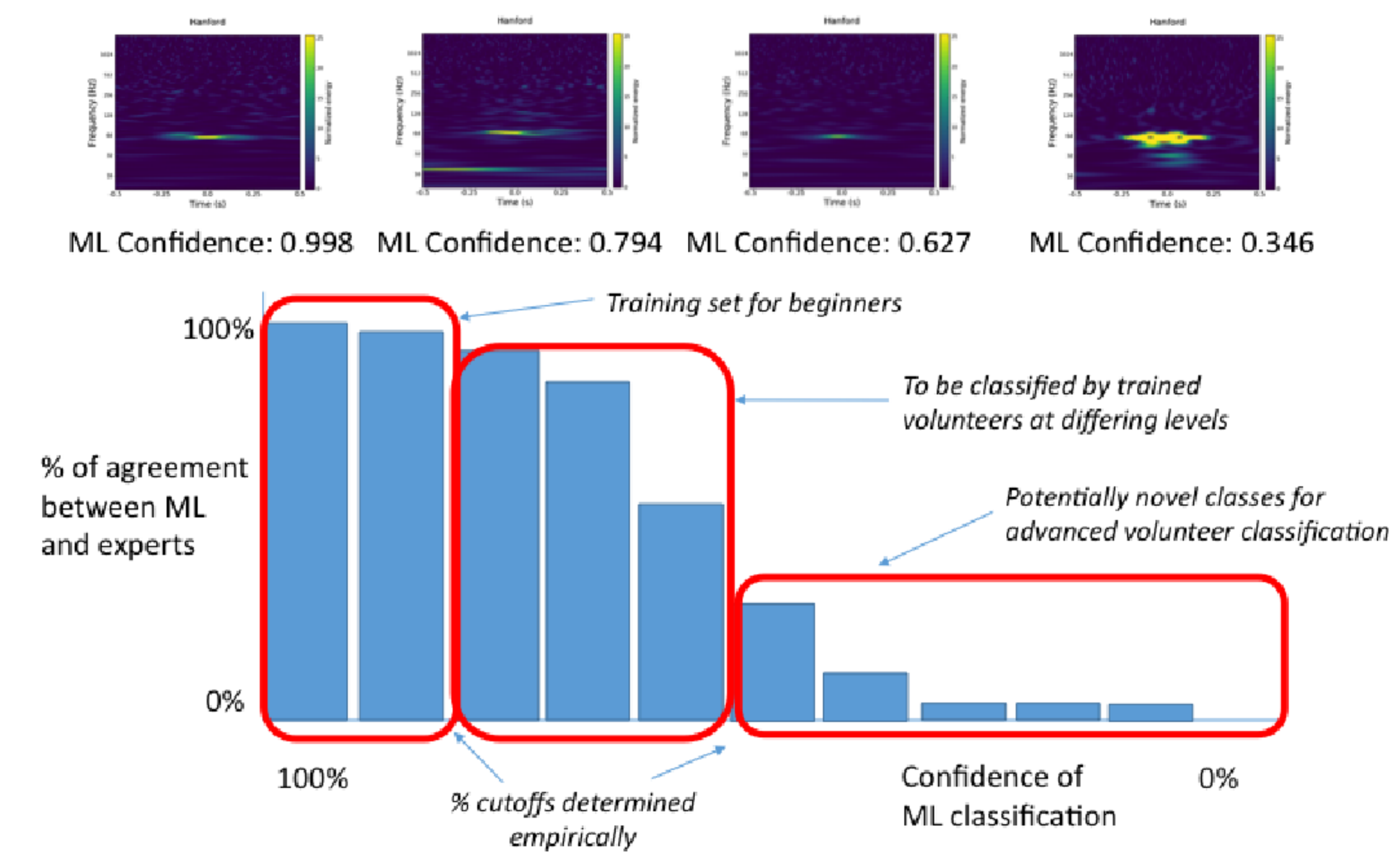}
\caption{Relationship between machine learning confidence in glitch classification (x-axis) and proportion of images from that class assessed by human volunteers at different skill levels. Example glitches classified as a single class (``Power Line" glitches) with differing machine learning confidence scores are shown above.}
\label{fig:confidence}
\end{figure}

Glitches are first sent through the machine learning classifier, which is trained on a set of images pre-classified by experts (see section \ref{subsec:training_sets}) and images retired from the project. Based on the machine learning confidence of the classification of each image, it is routed either to beginning, intermediate, or advanced workflows, as illustrated in Figure \ref{fig:confidence}. 

Similarly, based on their expertise and reliability level as determined by their performance in classifying (described in section \ref{subsec:crowdsource_classifier}), volunteers are divided into three levels that correspond to the beginning, intermediate, and advanced workflows. Through the Gravity Spy interface, LIGO detector characterization experts will be fed glitches for which the most advanced users cannot reach a consensus. Each volunteer starts at the simplest level and can be promoted to higher levels based on their performance.

As images are classified, the models of both the image and the volunteer are updated. The destination of a glitch (whether it stays in its current workflow, moves to a more difficult workflow, or is retired) is determined by a combination of machine learning and user confidence posteriors. If an image achieves high enough confidence in its classifications, it will be retired and added to the training set to further improve the performance of the machine learning classifer. 

The system is built to optimize the retirement of images. Most citizen science projects rely solely on number of classifications as a gauge for retirement (e.g. any image that has 20 classifications is retired from the project). However, this methodology presents multiple problems. Images can be retired even when there is strong disagreement on the correct class. Furthermore, many classifications are essentially wasted on easy images, which may only require a few identical classifications for accurate retirement, whereas difficult images that require deeper analysis may not receive enough classifications. By relying on the combination of machine learning and user classification, and weighting user classifications differently based on their prior performance, the Gravity Spy project aims to ameliorate such issues. 

\subsection{Machine Learning}

The following section describes the application of machine learning to the problem of classifying images in the Gravity Spy system and how the classifications contributed by volunteers are used to update models of both machine learning image classification and volunteer capabilities. 

\subsubsection{Image Classifier}

\begin{figure}[b]
\centering
\includegraphics[width=0.8\textwidth]{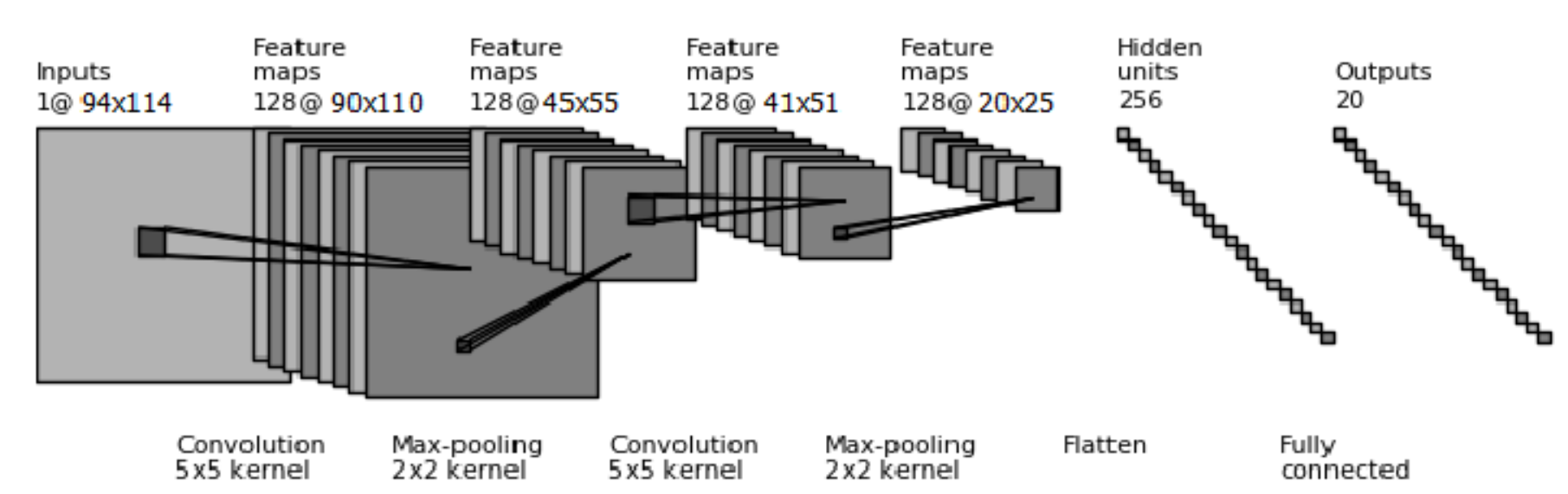}
\caption{Deep CNN used for glitch image classification. The network has been introduced on top of the four merged glitch durations. Dimensions of the kernels and feature maps are in units of pixels.}
\label{fig:conv_NN}
\end{figure}

Deep learning is a branch of machine learning which utilizes algorithms that attempt to model high level abstractions in data by using multiple processing layers, composed of multiple linear and non-linear transformations. The Gravity Spy system uses a deep model with Convolutional Neural Network (CNN) layers, which has shown great performance and is considered the state-of-the-art in image classification \cite{imagenet}. 
Another reason for exploiting deep learning is its scalability; compared to traditional machine learning methods such as support vector machines (SVMs), deep learning can handle and take advantage of copious amounts of data. Figure \ref{fig:conv_NN} illustrates the machine learning process used. 


Many studies (e.g. \cite{saraieee,nedaeu}) have shown that using multiple sources of information can improve the overall performance of classification. In this project, the multiple glitch durations that are also shown to Zooniverse volunteers are utilized. These durations are merged into a square form so that kernels can slide over all different durations and learn the glitch patterns. Two convolution layers are utilized first. The kernels slide over the input matrix, multiplying their corresponding weights to the input matrix and outputting a new matrix. The output of each kernel is known as a \textit{feature map}. 

Feature maps are usually subsampled using a max (or mean) operation. Here, max-pooling is used for down sampling - a square matrix slides over the feature map and gives the maximum value among the elements inside it. A layer of activation functions is used to determine the output of a given neuron. The Gravity Spy model uses a popular activation function known as rectified linear unit (ReLU) which is defined as $\textrm{max}(0,x)$. Then, a fully connected layer is applied. Each node in the fully connected layer is connected to all nodes of the previous layer.

The final layer is a softmax layer with $20$ outputs. Softmax is an fully connected layer with the same number of nodes as the number of classes, and is widely used as the final layer in multi-class classification tasks. The output of the softmax layer, when image ``$i$" is given as the input to the classifier, is defined as 
\begin{equation}
o_i^c = \frac{e^{w_c^Tx}}{\sum_{c=1}^{C}e^{w_c^Tx}} \qquad  \textrm{  for  }c=1,\cdots, C
\label{eq:cros1}
\end{equation}
\noindent
where $o_i^c$ is the output score of class $c$ for glitch $i$, $x$ is the output of the layer before softmax when image ``$i$'' has been given as input to the model, and $w_c$ is the vector of weights connecting the output of the previous layer to $c^\textrm{\textit{th}}$ node in softmax layer. $C$ represents the total number of classes, in our current case $20$. The output score of the softmax layer, $o_i^c$, is used as the probability distribution found by the image classifier. The score vector obtained from machine learning for image $i$ is defined as follows:
\begin{equation}
\mathbf{p}^{ML}_i = [o_i^1, \cdots,o_i^c,\cdots,o_i^C]
\end{equation}

The next step is to train the model. The model optimizes a loss function defined on the training data, using cross-entropy: 

\begin{equation}
\textrm{loss} = -\sum_{j=1}^{N} \sum_{c=1}^{C} y_j^c \log o_j^c
\label{eq:cros}
\end{equation}
\noindent
where $o_j^c$ is the model's output for class $c$ when the $j^{\textrm{\textit{th}}}$ training sample is given to the network, $y_j^c$ is equal to unity if the $j^{\textrm{\textit{th}}}$ sample is from class $c$, otherwise it is zero, and $N$ and $C$ are the total numbers of the training samples and classes, respectively. To optimize the objective function, the Adadelta \cite{zeiler2012adadelta} optimizer is used. This optimizer monotonically decreases the learning rate and shows good performance in our experiments. More details about the proposed machine learning image classifier and experiments can be found in \cite{saraicassp}.

\subsubsection{Crowdsource Classifier}\label{subsec:crowdsource_classifier}

As noted above, the system will maintain a model of each volunteer’s ability to classify glitches of each class and will update the models after each classification (e.g. increasing its estimate of the volunteer's ability when they agree with an assessment and decreasing it if they disagree). When the volunteer model shows that a volunteer’s abilities is above a certain threshold, the volunteer will be advanced to the next workflow level, in which they will be presented with new classes of glitches and/or glitches with lower machine learning confidence scores. In addition, the movement of images through the project is determined by these volunteer performance models, as well machine learning and volunteer classification. As a collective, these algorithms are referred to as the \textit{crowdsourcing classifier}. Further details regarding the crowdsourcing classifier will be presented in an upcoming publication \cite{emre}. 

A confusion matrix is assigned to each volunteer to record their labeling performance. It is defined as $\mathcal{M}^k \in {\mathbb N}^{C \times C}$ for the $k^\text{th}$ volunteer, where $C$ denotes the total number of classes. An entry of this matrix, $m_{pq}^k$ gives the number of samples belonging to class $p$ labeled as belonging to class $q$ by the $k^{th}$ volunteer. All entries will be initiated as 0 and updated when an image from the golden set is labeled by the volunteer. It will also be retrospectively updated with the labels of testing images that are retired.

Using a volunteer's confusion matrix $\mathcal{M}^k$, a reliability measure is defined for volunteer $k$ as the vector $\mathbf{a}^k=[\alpha_1^k, \cdots, \alpha_c^k,\cdots, \alpha_C^k] \in {\mathbb R}^{C \times 1}$, where $\alpha_c^k$ quantifies the reliability of volunteer $k$ in classifying samples of class $c$. It is defined as: 

\begin{align}
&\alpha_c^k =\frac{m_{cc}^k}{\sum_{j=1}^{C} m_{cj}^k} = p(\hat{y}^k=c|y=c) \qquad \textrm{for} \quad c \in \{1,\cdots,C\} 
\label{parameters}
\end{align}
\noindent
where $\alpha_c^k$ is also equal to the probability that the $k^{th}$ volunteer provides a label $\hat{y}^k$ for an image, as belonging to class $c$, given the true label $y$ is indeed equal to $c$.

After modeling the volunteers' reliability, the classification of a test sample of images using multiple annotations is determined. 
A test image is initially provided to the machine learning classifier which outputs a probability vector $\mathbf{p}^{ML}_i$.
Depending on this machine learning confidence, the test sample is forwarded to volunteers in a given workflow, who provide classification labels to this image. 
The developed algorithm uses the machine learning probabilities and volunteer classification labels to predict the true label \cite{emre}.

With the assigned labels from $R_i$ volunteers for a given image $i$, the goal is to fuse these labels and find the posterior probabilities $p(y_i^{cr}=j| \hat{y}_i^1,\cdots,\hat{y}_i^{R_i})$ for $j \in \{1, \cdots, C\}$, where ${y}^{cr}_i$ is the predicted label from crowdsourcing 
information. The final predicted label $\tilde{y}_i$ is calculated as:

\begin{equation}
\tilde{y}_i = \textrm{argmax}_{j}  \frac{{p({y}_i^{cr}=j| \hat{y}_i^1,\cdots,\hat{y}_i^{R_i})+\mathbf{p}^{ML}_{i}(j)}}{\sum_{j=1}^C{p({y}_i^{cr}=j| \hat{y}_i^1,\cdots,\hat{y}_i^{R_i})+\mathbf{p}^{ML}_{i}(j)}} \label{finalposterior} 
\end{equation}
\noindent
where $\mathbf{p}^{ML}_{i}(j)$ denotes the $j^{\text{th}}$ component of $\mathbf{p}^{ML}_{i}$. 

As classifications are made, the initial priors provided by machine learning are replaced by the posterior probability of each class, which contains both machine learning and volunteer classification information. 
The posterior probabilities continually update until an image is retired or the image receives a predefined maximum number of volunteer classifications and is moved to a higher workflow to be investigated by more advanced volunteers. To decide on the retirement of the test image, a threshold $t_j$ is defined per class based on the difficulty of classifying glitches in that class. The threshold vector can be thus defined as \textbf{t} = $[t_1, t_2, \cdots,  t_C]^T$. 


Having the posterior probabilities of all the classes from Eq. \ref{finalposterior} and putting them in a vector $\textbf{y}^i =[y_{1}^i, \cdots, y_{C}^i] \in  {\mathbb R}^C$, the posterior probability vector $\textbf{y}^i$ can be compared with threshold vector $\textbf{t}$. If the entry of $\textbf{y}^i$ that carries the highest posterior probability is greater than the corresponding entry of $\textbf{t}$, the image is retired with label $j$ for which $y_{j}^i \geq t_j$. Then this retired image is sent to training set with label $j$ as its true label. If no entry of $\textbf{y}^i$ is greater than the corresponding entry of $\textbf{t}$, further action is needed. Based on the number of volunteers who have labeled the image, either more volunteers at the same level must label the image or the image is moved to a more advanced workflow.

As for volunteer promotion, when a volunteer labels images from the golden set, their confusion matrices are updated. Also, as test images are retired, the golden set is updated and the confusion matrices are updated retrospectively by comparing their labels with the label of the retired image. With Eq. \ref{parameters},  the $\mathbf{a}^k$ vector is calculated from the confusion matrix $\mathcal{M}^k$. Reliability threshold values are defined for each class: ($\textbf{T}_j = [T_1, \cdots, T_C]$). 
If all the values of the vector $\mathbf{a}^k$ exceed the threshold values in $\textbf{T}_j$, the volunteer is promoted to the next level. If not, they will need to do more correct classifications to be promoted. 

\subsection{Socio-Computational Research Support}

Finally, the socio-computational research component (yellow boxes in Figure \ref{fig:bigblockdiagram}) will allow for systematic measurement and experimentation with the performance of project components. Our first planned experiment is to compare the performance of volunteers who have gone through the training process described above to the performance of those who start right away with the full set of classes for classification (i.e. the typical approach for citizen science projects). By doing so, one can test if users who go through the training regimen contribute more and show better performance on the classification tasks. 

Second, the training system described above has a large number of parameters (e.g. how many and which classes to introduce at each level and the class-specific machine learning certainty cutoffs for images to be placed in each level). Experimentation will be useful to determine the optimal settings. For example, one can test the benefits and tradeoffs of advancing volunteers to higher levels more rapidly: quicker advancement might be good for motivation but negative for performance (and vice versa). 

Finally, the system will enable us to experiment with other factors that affect volunteer performance, such as the kinds of motivational messages provided or information on the novelty of glitches. A particularly interesting set of questions gauge the effects of feedback that can be provided to volunteers based on machine learning classification confidence. Again, it is possible that there are tradeoffs involved: letting a volunteer know the machine learning confidence score of an image might be useful feedback to improve performance but also potentially demotivating if the machine learning and the volunteer disagree, or if it leads to volunteers feeling that their contributions are unnecessary.

There are many unanswered questions about how volunteers will learn in this setting that go beyond the specifics of glitch classification. In particular is the concern of how much the volunteers will need to know about gravitational-wave astrophysics and the workings of the detectors that produce the glitches. Included as part of the workflows is a mini-course on gravitational-wave astrophysics and LIGO detector characterization that presents the next slide of the course after a given number of classifications. Additionally, there are background information pages on the site that describe the detector in more detail. Though the background pages are optional and one can opt-out of the mini-course, one can track which volunteers visit these pages to examine the impact on performance. Further details on the socio-computational research related to Gravity Spy can be found in \cite{kevin_conference}.

\section{Preliminary Results}\label{sec:results}

The full public launch of the Gravity Spy project was on October 12 2016, about a month before the planned commencement of LIGO's second observing run (O2). Through the initial renditions of machine learning models and beta-testing of the human interface, the preceding phases of this project have already shown promise in achieving high-level, multi-class glitch classification using true (rather than synthesized) LIGO detector data and the ability of the public to distinguish new categories of glitches. 

\subsection{Initial Machine Learning Performance}

As discussed in Section \ref{subsec:training_sets}, the initial machine learning training set consists of $7718$ total glitches from $20$ classes, using $75\%$, $12.5\%$, and, $12.5\%$ of the full set as training, validation, and test sets, respectively. The number of iterations and the batch size were set to $130$ and $30$, respectively. The classification of testing data achieved an average accuracy of $97.1\%$. 


\begin{figure}[t]
\centering
\makebox[\textwidth][c]{\includegraphics[width=1.2\textwidth]{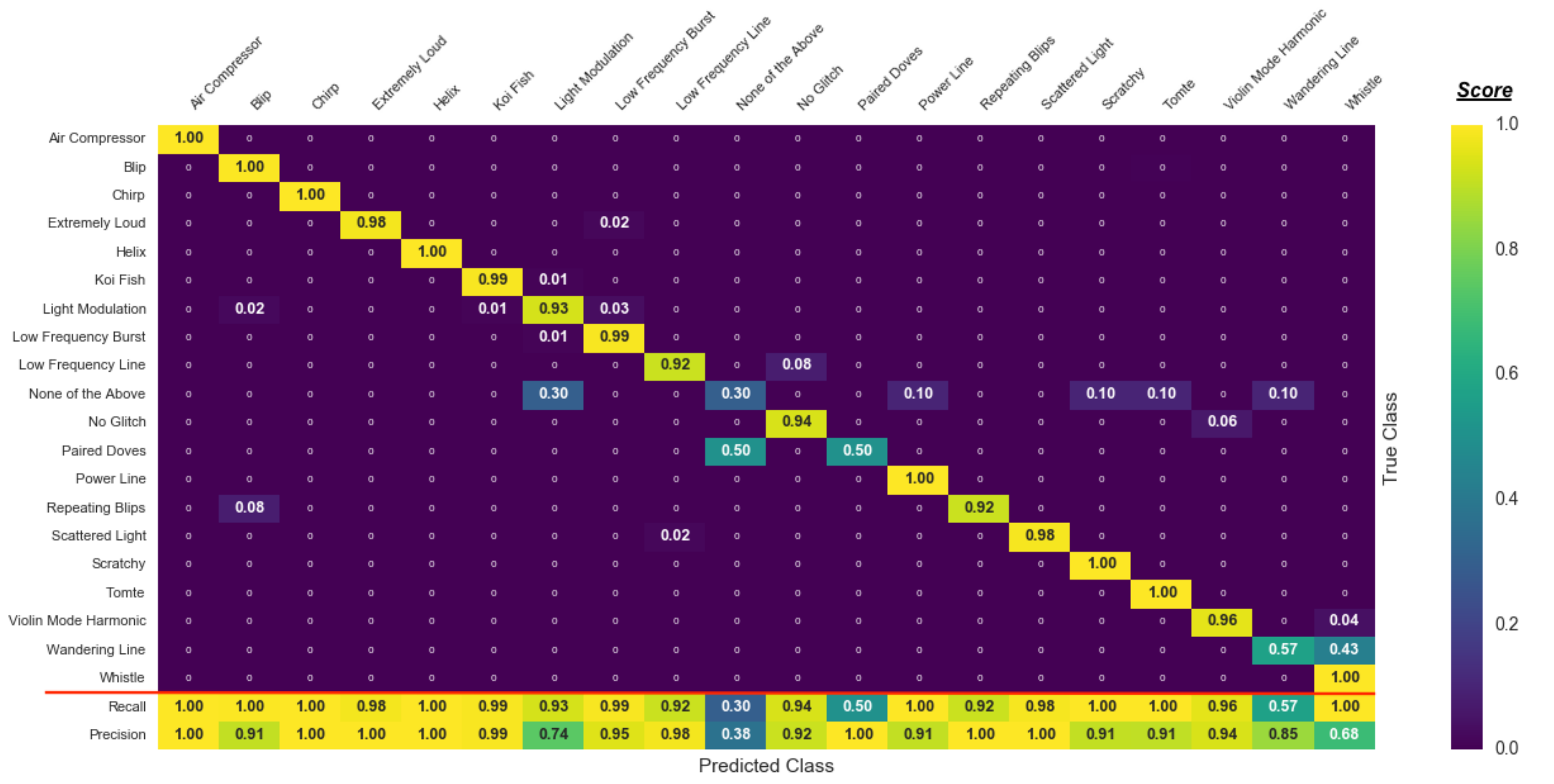}}
\caption{Confusion matrix for the 20 glitch classes in the testing set classified using CNNs, with recall and precision values appended below for reference. The x and y axes represent the predicted and true classes, respectively, and the confusion matrix is normalized by the total number of glitches in each class in the training set. Due to the normalization chosen, the diagonal elements are identical to the recall values for each class. Closer to unity in precision and recall values corresponds to a more accurate classification for a particular class.}
\label{fig:conf_heat_map}
\end{figure}

As can be seen in the training set breakdown (Table \ref{tbl:class_numbers}), the distribution of samples over classes is highly imbalanced. Therefore, it is better to study precision and recall values of each class to analyze the performance of the glitch classifier. \textit{Precision} is defined as the number of glitches that are correctly labeled as a particular class divided by the total number of glitches that are predicted as that particular class, gauging how often a classifier is correct when it predicts a glitch is in a given class. \textit{Recall}, also known as sensitivity, is the number of glitches predicted correctly as a particular class divided by the actual number of glitches in that particular class, in essence a measure of how often a classifier predicts a glitch in a particular class when it is actually in that class. These values are presented in Figure \ref{fig:conf_heat_map}. 

As one can observe from Figure \ref{fig:conf_heat_map}, the precision and recall values are near unity for most classes. Certain classes, particularly classes that suffered from a low number of training samples (e.g. ``Wandering Line" and ``Paired Doves") or a high variability in morphological characteristics (e.g. ``None of the Above" and ``No Glitch"), achieved lower precision and recall values. ``None of the Above" and ``No Glitch" are not defined by specific morphological traits. ``None of the Above" is the category which harbors all glitches that do not fit in the other $19$ classes. Therefore, this class does not have a specific morphological distribution over sample space. The ``No Glitch'' category has a similar property, as this class consists of all glitches which do not have intense energy in the image, and the low-level noise does not have a consistent morphology through the training set. Though not morphologically defined compared to the other classes, the inclusion of these two \textit{catch-all} classes allows for the full classification of the dataset, and provides a medium for determining new classes of glitches as the project progresses. The challenge of the classification of ``Paired Doves'' and ``Wandering Line" groups is likely due to a lack of samples, as these two classes have the lowest number of samples with $30$ and $44$, respectively. 

\subsection{Gravity Spy System Beta Testing Results}\label{subsec:beta}

The Gravity Spy project launched three Beta versions to test the user interface and user promotion in April, June, and September 2016, each of which lasted approximately one week. During this time, a version of the project was made public and promoted to a small subset ($\sim$2000) of Zooniverse volunteers. The main goal of the beta testing was to check the functionality of the site and to receive feedback on the interface design. However, the activity on the site also proved the basic premise of the project: volunteers can reliably classify glitches and identify new morphological classes. Beta testing of the website engaged over 1400 users and delivered over 45,000 glitch classifications. This activity in turn led to hundreds of conversation threads on the website’s talk forum and fostered excitement and intrigue for the nascent field of gravitational-wave astrophysics. The work culminated in the discovery of multiple new and substantial glitch categories from LIGO’s first observing run, including glitches which would later receive the names ``Paired Doves" \cite{paired_doves} and ``Helix" \cite{helix}. In particular, the discovery of the ``Paired Doves" class proved significant in LIGO detector characterization endeavors, as this glitch resembles signals from compact binary inspirals and is therefore detrimental to the search for such astrophysical signals in LIGO data. The project activity during the Beta versions is testament to the ability of citizen science projects to engage and involve the public in scientific advancement. A deeper analysis of these morphologies with regard to LIGO detector characterization and further techniques to optimize the integration of citizen science output to large-scale data analysis will be presented in future publications \cite{future_O1_pub,future_scotty_pub}.

\begin{figure}[t]
\centering
\includegraphics[width=3.in]{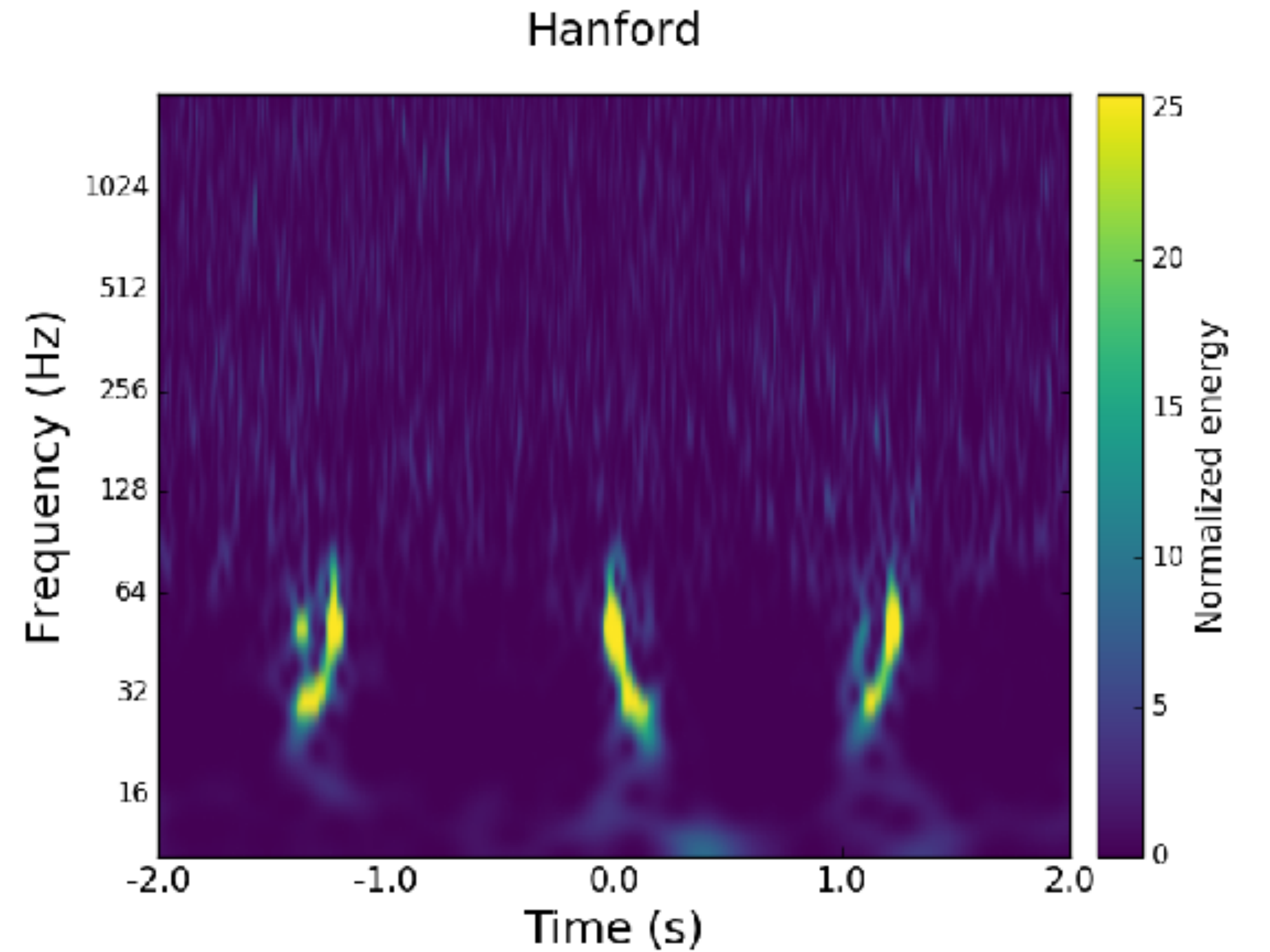}
\includegraphics[width=3.in]{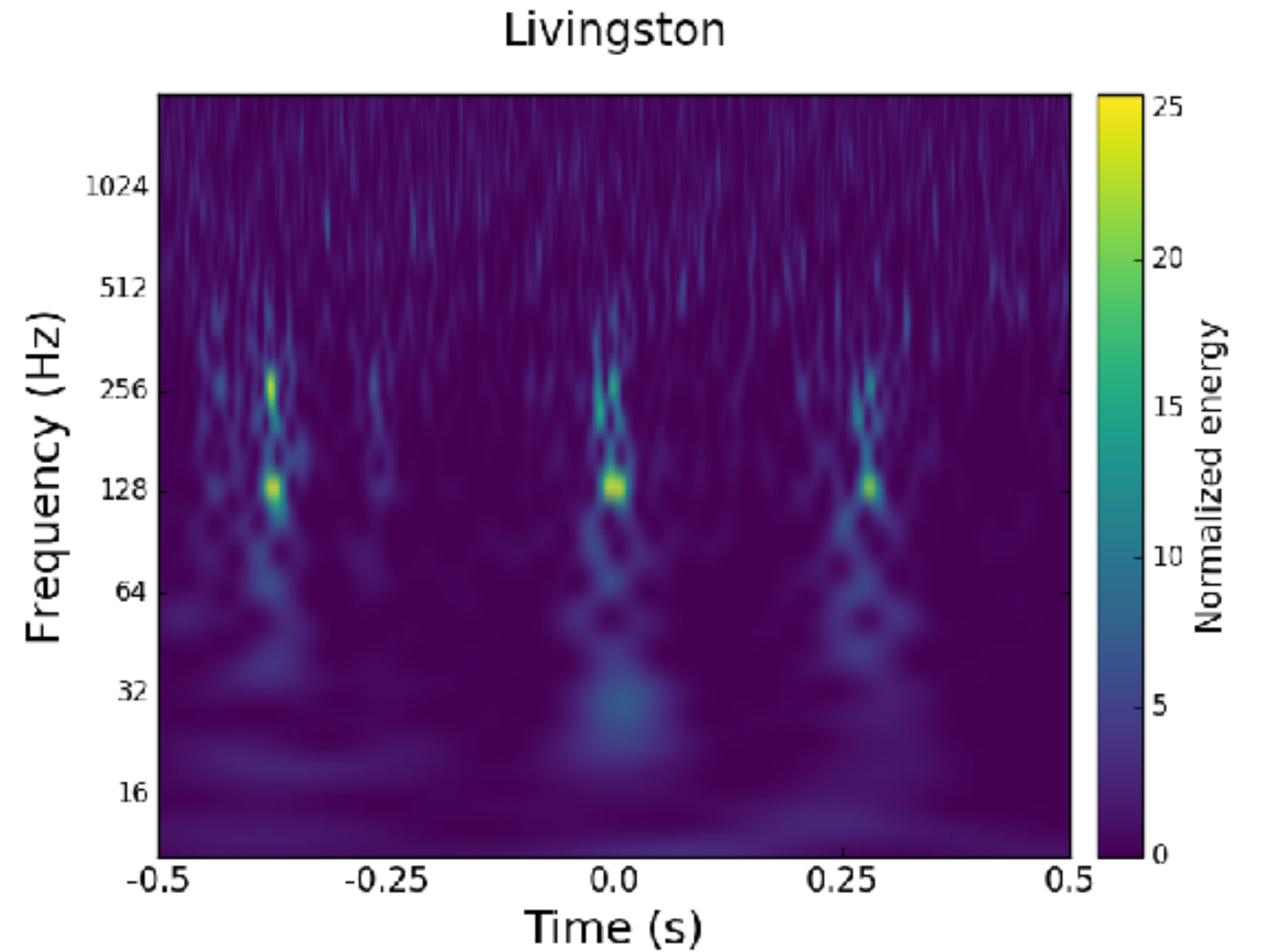}
\caption{Two new O1 glitch classes uncovered during Gravity Spy beta testing: ``Paired Doves" (left) and ``Helix" (right). ``Paired Doves" \cite{paired_doves} resemble chirps, but alternate between increasing frequency and decreasing frequency. These glitches are potentially related to 0.4 Hz motion of the beamsplitter at the Hanford detector. ``Helix" \cite{helix} are possibly related to glitches in the auxiliary lasers (called photon calibrators) that are used to push the LIGO mirrors and calibrate the detectors.}
\label{fig:beta_glitches}
\end{figure}

\section{Conclusions and Future Prospects}\label{sec:future}

As LIGO searches for gravitational waves, the Gravity Spy project will endeavor to improve the understanding of the LIGO detectors and reduce the impact of harmful noise, all while engaging the general public in gravitational-wave physics. The full launch of the Gravity Spy project on October 12 2016 incorporated the machine learning analysis and crowdsource classifier into the system, providing each user with a tailored progression through the multiple workflows and pairing machine learning confidence scores with user classifications to optimize the retirement of images and classification accuracy. The project shows clear utility in aiding LIGO detector characterization and creates an avenue to analyze the socio-computational interaction. 

Each day during LIGO's upcoming observing runs, the Gravity Spy system will generate Omega Scans of triggers that have passed low-latency data quality cuts and fit within the SNR and frequency thresholds defined in Section \ref{sec:dataprep}. These newly-acquired images will be analyzed using the most current renditions of the machine learning classifier, and integrated into the testing sets available for human classification. As images are retired from the test set, they are added to the machine learning training sets, which re-trains whenever 100 new images are retired and appended. Daily pages summarizing the results are available to all LSC members. 

When new classes appear in the detector and trends in the ``None of the Above" class emerge (via clustering of descriptive features from the follow-up questions and collections on the Gravity Spy Talk forum), new categories are added to the interface at the discretion of the Gravity Spy team. By doing so, the project maintains the ability to evolve with the detectors. In addition, the data synthesis for this project can adapt to the activity of the users; adjusting the SNR threshold of triggers will greatly affect the number of glitches that are generated from the LIGO data stream, and lowering this threshold will provide many more difficult images for users to analyze. 

As the project progresses, continual engagement of volunteers will be cultivated by providing complementary data and new tools to aid in the classification (e.g. the ability to view spectrograms from auxiliary channels of data, deeper classifications that included sub-classes of morphologies, and tools to support the discovery of new glitch classes and collaboration among volunteers). This, along with continued interaction between project scientists and volunteers on the Talk forum, will foster sustained engagement in the project. Gravity Spy also presents a test bed for socio-computational interaction. Some of the many possible empirical tests that will be implemented include presenting a different interface to subsets of users to examine its impact on user activity (e.g. retracting the training regimen, changing the wording of the project pitch) and analyzing the classification output to investigate how users learn (e.g. examining if the use of filters diminishes for a user over time, inspecting the performance of a user over time). Furthermore, as the human and machine learning components of the project utilize the exact same data for their classification endeavors, it will provide an interesting comparison of each classifier on a level playing field. 

Though crowdsourcing models have proven effective in data analysis endeavors across multiple scientific disciplines, the exponential growth of data acquisition necessitates a smarter way to perform citizen science. The sheer amount of data that modern projects produce will soon outstrip human volunteer time, and simple crowdsourcing methods will no longer suffice as a means to scrutinize such sets. The coupling of citizen science to machine learning algorithms that resourcefully choose the optimal data for human classification is essential to preserve crowdsourcing as a powerful means of data analysis. The integration of human and computer classification schemes will maintain citizen science as a prolific scientific tool and allow it to scale with the ever-increasing datasets of the future. 

\section*{Acknowledgments}
The Gravity Spy team would like to acknowledge and thank the many Zooniverse volunteers who provided invaluable feedback during Gravity Spy beta tests, and delivered initial glitch classifications for which to test the methods presented in this paper. In addition, the team would like to thank the Detector Characterization working group of the LSC for useful comments and suggestions, in particular Jess McIver for input during the planning phases, Chris Pankow for useful discussions, and Duncan Macleod for technical support and thorough comments on this manuscript. We would like to thank our undergraduate research team: Luke Calian, Jessie Duncan, Ethan Marx, Isa Patane, Leah Perri, and Ben Sandeen, for contributing to multiple components of the project, including the building and curation of the initial training set of glitches. Gravity Spy is partly supported by the National Science Foundation, award INSPIRE 15-47880. This paper has been assigned LIGO document number ligo-P1600303.

\section*{References}
\bibliographystyle{iopart-num}
\bibliography{GravitySpy.bib}

\end{document}

%% file: definitions.tex
\usepackage{bm}
\usepackage{color}
\usepackage{amssymb}

\def\bi{{\mathbf i}}
